\documentclass[fleqn,10pt]{wlscirep}
\usepackage{graphicx}
\usepackage{dcolumn}
\usepackage{bm}
\usepackage{float}
\usepackage{amsmath}
\usepackage{upgreek}
\usepackage{multirow}
\usepackage{color}
\usepackage{cases}
\usepackage{epstopdf}
\usepackage[english]{babel}
\usepackage{csquotes}
\title{Stable attosecond electron bunches from a nanofiber driven by Laguerre-Gaussian lasers}

\author[1]{Li-Xiang Hu}
\author[1,2,*]{Tong-Pu Yu}
\author[2,3,4]{Zheng-Ming Sheng}
\author[5]{Jorge Vieira}
\author[1]{De-Bin Zou}
\author[1,6]{Yan Yin}
\author[2]{Paul McKenna}
\author[1]{Fu-Qiu Shao}

\affil[1]{College of Science, National University of Defense Technology, Changsha, 410073, China}
\affil[2]{SUPA Department of Physics, University of Strathclyde, Glasgow G4 0NG, UK}
\affil[3]{Collaborative Innovation Center of IFSA (CICIFSA), Key Laboratory for Laser Plasmas (MoE) and School of Physics and Astronomy, Shanghai Jiao Tong University, Shanghai, 200240, China}
\affil[4]{Tsung-Dao Lee Institute, Shanghai, 200240, China}
\affil[5]{GoLP/Instituto de Plasmas e Fus\~ao Nuclear, Instituto Superior T\'ecnico, Universidade de Lisboa, 1049-001 Lisbon, Portugal}
\affil[6]{Institute of Applied Physics and Computational Mathematics, Beijing, 100094, China}

\affil[*]{Corresponding author: tongpu@nudt.edu.cn}

\begin{abstract}
Generation of attosecond bunches of energetic electrons offers significant potential from ultrafast physics to novel radiation sources. However, it is still a great challenge to stably produce such electron beams with lasers, since the typical sub-femtosecond electron bunches from laser-plasma interactions either carry low beam charge, or propagate for only several tens of femtoseconds. Here we propose an all-optical scheme for generating dense attosecond electron bunches via the interaction of an intense Laguerre-Gaussian (LG) laser pulse with a nanofiber. The stable bunch train results from the unique field structure of a circularly polarized LG laser pulse, enabling each bunch to be phase-locked and accelerated forward with low divergence, high beam charge and large beam-angular-momentum. This paves the way for wide applications in various fields, e.g., ultrabrilliant attosecond x/$\gamma$-ray emission.
\end{abstract}
\begin{document}

\flushbottom
\maketitle
%
%
\thispagestyle{empty}

\section*{Introduction}

The generation of ultrashort intense laser pulses pushes the laser-matter interaction to the relativistic regime where the relativistic electron dynamics in the laser fields become dominant. Optical processes driven by relativistic particle motion in plasmas,
so-called “relativistic plasma optics”, enable investigation of ultrafast physics phenomena and the development of compact radiation sources~\cite{Mourou06}. The short energetic electron beam has significant potential for applications such as electron diffraction and microscopy~\cite{McMorran11}, electron 4D imaging~\cite{Baum07}, and electron injection into a free electron laser (FEL). Especially, this leads to the production of ultrashort x-ray radiation sources with duration down to the attosecond level~\cite{krausz}, where femtosecond and even attosecond relativistic electron bunches with a narrow energy spread, small divergence angle and large flux are of crucial importance~\cite{Corde13}. For this purpose, significant efforts have been dedicated to acquiring high-quality ultrashort electron bunches using the normal Gaussian laser pulses, for example, by laser irradiating high density plasma boundaries of a channel, hollow cone, wire or droplet target~\cite{Naumova04,Ma06POP,Karmakar2007,Liseykina10,YTP13PRL,Hu16JAP,YTP14APL}, underdense plasmas~\cite{Luttikhof10,LFY13} or nanofilms~\cite{Wu10,Kulagin07}, by the inverse FEL process~\cite{Sears08}, direct laser acceleration in vacuum~\cite{Stupakov01}, or by a relativistic electron beam scattering off interfering laser waves~\cite{Dodin07}. In spite of these different concepts and schemes, it is still very challenging to produce stable and high-quality attosecond electron bunches for potential applications. Even with pC charges, the space-charge repulsion of electrons within the bunch severely limits the bunch lifetime to several tens of femtoseconds with either a low beam charge or large divergence angle. A new solution to suppress the space charge effect in the relativistic regime is required for stable dense electron bunch generation and acceleration.

A proper laser field structure in time and space potentially provides effective manipulation of atoms~\cite{Grier03} and charged particles~\cite{Vieira14,Wang15,Zhang16}, including the generation and acceleration of ultrashort electron bunches. For example, a Laguerre-Gaussian (LG) laser pulse is characterized by a hollow intensity distribution, a spiral equiphase surface, and orbital-angular-momentum (OAM)~\cite{Allen92}. The OAM transfer from the laser pulse to relativistic electrons is a topic of fundamental importance on relativistic optics~\cite{Haines2001}. It is shown theoretically and experimentally that a relativistic intense LG laser pulse can be generated using a spiral-shaped foil~\cite{Shi14,Denoeud17}, via stimulated Raman scattering~\cite{Vieira16NC} or plasma holograms~\cite{Leblanc17}. This may offer a new approach to generate and manipulate electron bunches in micron scale by relativistic LG-mode laser-plasma interaction.

In this paper, we report an innovative scheme to generate a dense and stable attosecond electron bunch train efficiently from a laser-driven nanofiber target. In the scheme, a circularly polarized LG laser pulse with OAM but without total AM is employed. The attosecond electron bunches are first dragged out by the radial laser electric fields and are quickly phased-locked and accelerated forward by the longitudinal electric fields. The stable attosecond bunch train results from the unique field structure of the circularly polarized LG laser pulse, enabling each bunch to converge into a dense disk and propagate with a narrow energy spread, high beam charge and large beam-angular-momentum (BAM). Full three-dimensional (3D) particle-in-cell (PIC) simulations indicate that the stable attosecond structure remains intact for several hundreds of femtoseconds, paving the way for many potential applications in various fields.
\begin{figure}
\centering
\includegraphics[width=9.2cm]{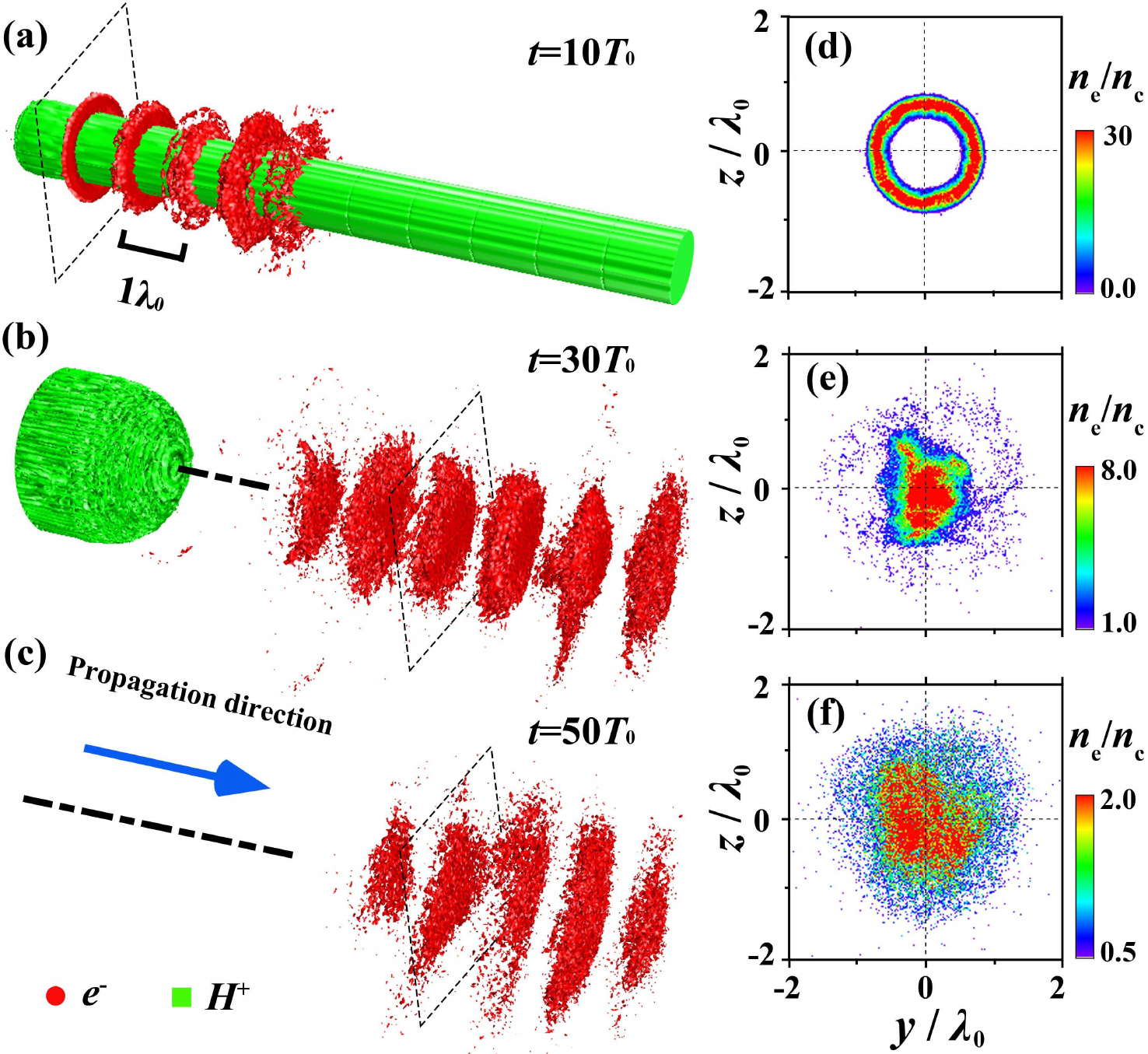}%
\caption{\label{f1}
Density evolution of the attosecond electron bunches (a-c). The panels (d-f) show the corresponding transverse density distribution of a bunch as marked in the panel (a-c).}
\end{figure}

\section*{Results}

Figure~\ref{f1} presents the typical simulation results obtained with $a_0=30$. At $t=10T_{0}$, a series of short electron bunches with a 1$\lambda_{0}$ spacing are generated along the nanofiber. Each bunch is isolated and doughnut-like, which is markedly different from the sawtooth-like or helical structure in a plasma channel, a hollow cone, a wire, or in the case of micro-droplets using normal Gaussian laser pulses~\cite{Naumova04,Liseykina10,Hu16JAP,Ma06POP,YTP13PRL,YTP14APL}. These electron bunches move almost at the speed of light with a longitudinal thickness corresponding to a fraction of the laser wavelength ($\sim0.2\lambda_0$, i.e., $\sim660$ as in our simulations at $t=10T_0$). The inner ring diameter is only 1.0$\lambda_0$, with the maximal density of $30n_c$. When the annular electron bunches leave the nanofiber, they are further accelerated and each bunch converges into a dense disk with beam charge up to 0.3$~$nC, instead of being dispersed in space as in the case with normal Gaussian laser pulses~\cite{Hu16JAP,Naumova04,Ma06POP,YTP13PRL,Liseykina10,YTP14APL}. As time progresses, the thickness of the disk decreases to, $\sim200$ as, while its peak density is still up to 2$n_c$ at $t=50T_{0}$. The attosecond structure is very stable and remains intact for a duration $>300~$fs (see Supplementary Fig. S1 online), which is about one order of magnitude longer than in other cases~\cite{Naumova04,Ma06POP,Liseykina10,YTP13PRL,Hu16JAP,YTP14APL,Kulagin07}.

\begin{figure}\suppressfloats
\centering
\includegraphics[width=8.4cm]{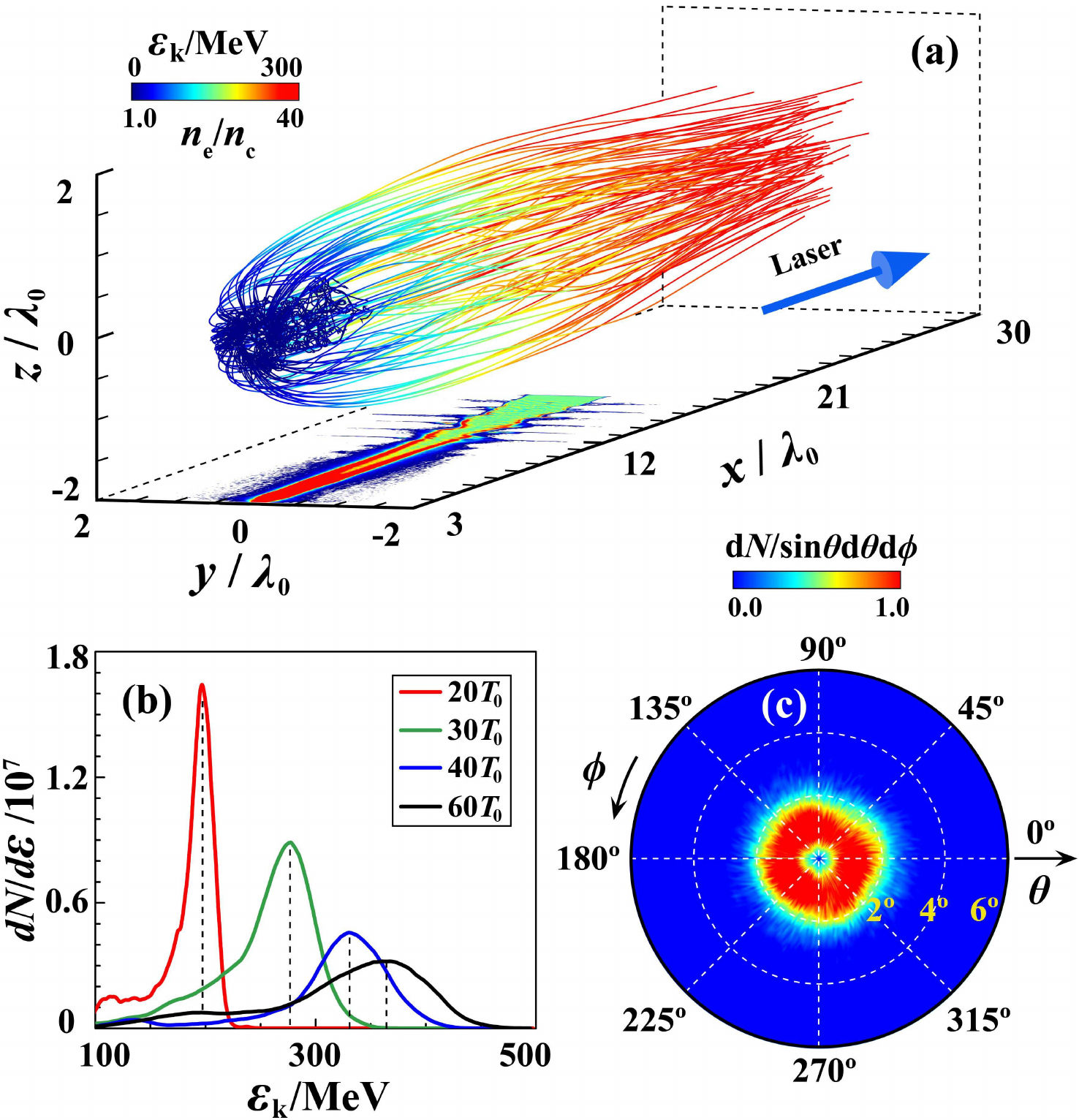}%
\caption{\label{f3}
Trajectories of some energetic electrons until $t=35T_0$ (a), the energy spectrum of the bunch (b) as marked in Fig. 1, and the electron divergence at $t=60T_0$ (c). Here, $\theta=\arctan(p_{\bot}/p_x)$, $\phi=\arctan(z/y)$, and $p_{\bot}=({p_y^2+p_z^2})^{1/2}$. The bottom panel in (a) shows the electron density distribution in the $xy$ plane at $t=20T_0$.}
\end{figure}

The formation of annular attosecond electron bunches in our scheme results from the unique field structure of the circularly polarized LG laser pulse. Here we considered a left-hand circularly-polarized LG-mode laser pulse with an OAM per photon but without total AM, which is characterized by simultaneously cylindrically-symmetric radially-polarized and azimuthally-polarized components. For convenience, we first convert the laser transverse electric fields into radial and azimuthal equivalents in the cylindrical coordinate system, owing to $E_r$=$\cos \varphi  \cdot E_y$+$\sin \varphi  \cdot E_z$ and $E_{\varphi}$=$-\sin \varphi  \cdot E_y$+$\cos \varphi  \cdot E_z$:
\begin{eqnarray}
\left(
  \begin{array}{c}
    E_{r} \\
    E_{\varphi} \\
  \end{array}
\right)=\frac{\sqrt{2e} E_{L0}r \sigma _0}{\sigma ^2} e^{-\frac{r^2}{\sigma^2}} g\big( x-ct \big)\left(
  \begin{array}{cccc}
   \sin\psi^{'}\\
    -\cos\psi^{'}\\
  \end{array}
\right),~\label{eq1}
\end{eqnarray}

\noindent where $\psi^{'}=-k[x+r^2/(2R)]+\omega_{0} t-2\arctan(x/f)$ is the transformed phase term. One can see that both the electric field $E_r$  and $E_{\varphi}$ are circularly-symmetric in the transverse cross-section and have nothing to do with the azimuthal angle $\varphi$. It is distinct from the normal radially polarized laser pulses~\cite{Karmakar2007,Salamin2007,Shell14,Zam2017}, where only the radial component exists for the electric field. This unique optical field structure would benefit the dense electron bunch acceleration and propagation greatly. When $E_r$ is positive, the electrons in the skin layer of the fiber are compressed, forming a rippled structure on the fiber surface (see the bottom panel in Fig.~\ref{f3}(a)); as it becomes negative, the electrons are subject to a radial outward force and are dragged out of the nanofiber, so that an attosecond bunch forms. Since $E_r$ is also periodic along the $x$ direction with a period of 1$T_0$, the spacing of the attosecond electron bunches is exactly 1$\lambda_{0}$, as seen in Fig.~\ref{f1}(a-c). However, when we change the laser handedness, the electric field structure varies accordingly. As a result, the attosecond structure disappears and two separate electron helices surrounding the nanofiber are observed. Thus the field struture of the LG laser pulse plays a key role in the electron dynamics.
\begin{figure}\suppressfloats
\includegraphics[width=9.2cm]{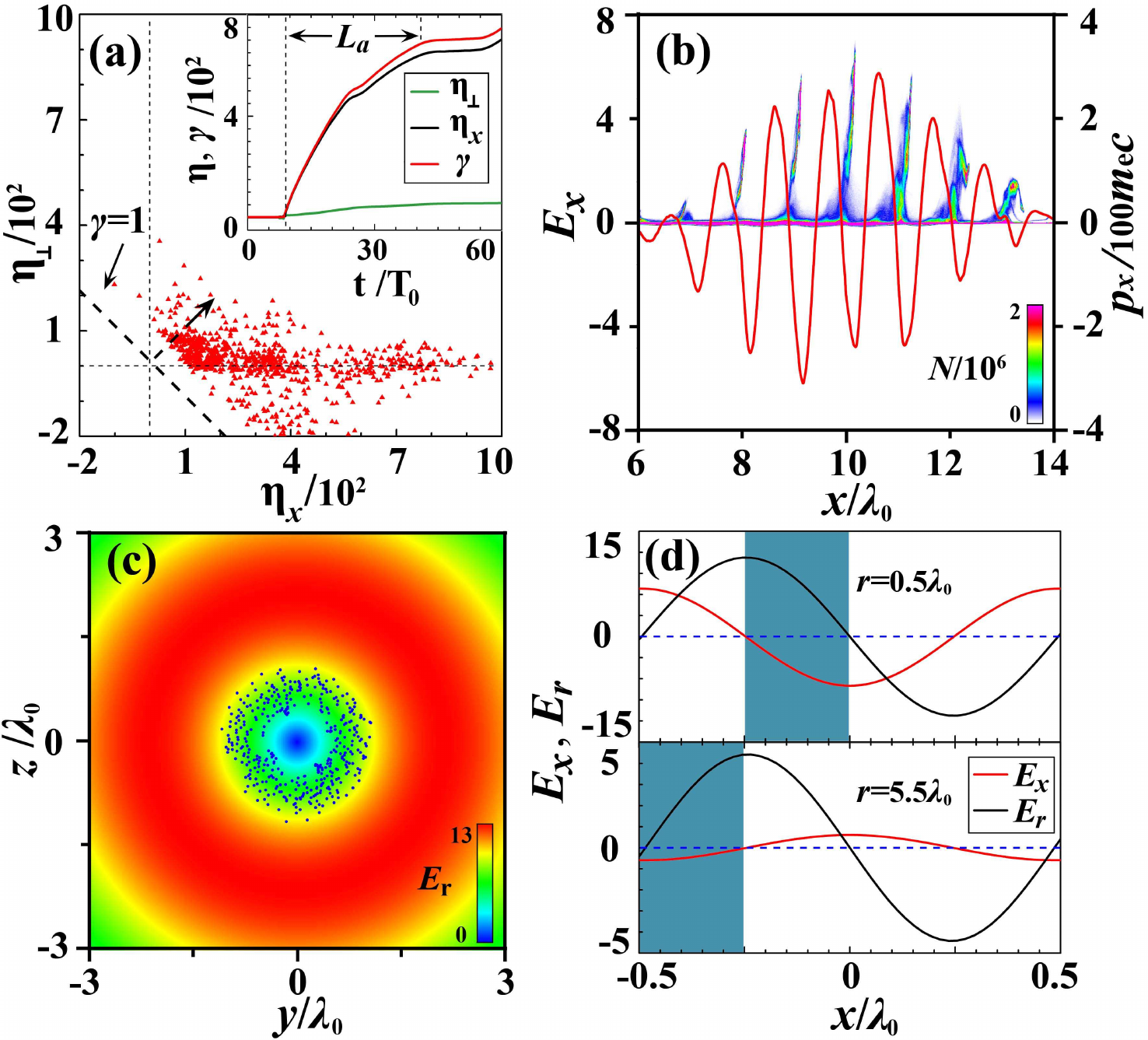}%
\centering
\caption{\label{f2}
The tracked electrons in $(\eta_{x},\eta_{\perp})$ (a) and ($x$, $p_x$) space (b) at $t=16T_{0}$. The red curve shows the axial electric field $E_{x}$ in (b). The radial electric field $E_r$ experienced by an electron bunch marked in Fig.~\ref{f1} at $t=16T_{0}$, with blue dots representing the positions of the bunch electrons (c). Phase-locking of electrons in a single laser period with $r=0.5\lambda_0$ (top) and $5.5\lambda_0$ (bottom) (d). Here, the shadow marks the phase-locking zone. The electric field is normalized by $E_0=m_ec\omega_0/q_e$.}
\end{figure}

Once the attosecond electron bunches are dragged out, they experience acceleration by the longitudinal electric fields, which exist for a tightly focused circularly-polarized LG laser pulse. Here one can obtain longitudinal electric field from the Poisson equation, i.e., $E_x (r, \varphi, x)=-(i/k)(\partial E_y/\partial y+\partial E_z/\partial z)$:
\begin{eqnarray}
\begin{aligned}
E_x=\frac{\sqrt{2e} E_{L0} \sigma_0 }{ k \sigma^2 }e^{ - \frac{r^2} {\sigma^2} } g\Big( x-ct \Big)
 \left [\Big( \frac{2r^2}{\sigma^2}-2 \Big) \cos\psi^{'} -\frac{kr^2}{R}\sin\psi^{'} \right ],
\label{eq2}
\end{aligned}
\end{eqnarray}
\noindent When $x=0$, $E_x\sim-2\sqrt{2e} E_{L0}/(k\sigma_0)$ at the center axis and vanishes at $r=\sigma_0$; For $r < \sigma_0$, $E_x < 0$, which is able to accelerate electrons. For each single electron dragged out from the fiber, the electron relativistic equations of motion in the LG laser fields together with the energy equation can be written as $\gamma m_{e0}c(d\bm{\beta}/dt)=q_e[\bm{\beta}(\bm{\beta}{\bm\cdot}\bm{E})-(\bm{E}+c\bm{\beta}\times \bm{B})]$. Here $\bm{\beta}$ is the normalized electron velocity by the speed of light, $\gamma$ is the electron relativistic factor, $\bm{E}$ and $\bm{B}$ are the laser electromagnetic fields. In cylindrical coordinates $ (\bm{r}, \bm{\varphi},\bm{x})$, the equations are rewritten as
\begin{eqnarray}
\begin{aligned}
\frac{d p_r}{dt}&=-q_{e}\Big(E_r + v_\varphi B_x - v_x B_\varphi\Big),\\
\frac{d p_\varphi}{dt}&=-q_{e}\Big(E_\varphi + v_x B_r - v_r B_x\Big),\\
\frac{d p_x}{dt}&=-q_{e}\Big(E_x + v_r B_\varphi - v_\varphi B_r\Big),\\
m_{e0}c^2\frac{d \gamma}{dt}&=-q_{e}\Big(v_{r} E_r  + v_{\varphi} E_\varphi + v_{x} E_x\Big).
\label{eq2-2}
\end{aligned}
\end{eqnarray}
\noindent Since the azimuthal electromagnetic fields exist for a circularly-polarized LG laser pulse, it becomes a challenge to find an analytical solution to Eq.~(\ref{eq2-2}) above for each electron in the bunches. In order to explore the detailed dynamics, we track some typical electrons in the PIC simulations and present their energy evolution and distribution in Fig.~\ref{f2}(a). Here we separate the two driving terms in the energy equation and thus the energy gain due to the longitudinal and transverse electric fields is $\eta_{x}=-q_{e}\int_{0}^{t} v_{x} E_{x}dt/(m_{e0}c^2)$ and $\eta_{\perp}=-q_{e}\int_{0}^{t} \vec{v}_{\perp} \cdot \vec{E}_{\perp} dt/(m_{e0}c^2)=-q_{e}\int_{0}^{t}(v_{r} E_r  + v_{\varphi} E_\varphi )dt/(m_{e0}c^2)$, respectively, with $\gamma (t) = \gamma (0) + \eta_x + \eta_{\perp}$. As expected, the contribution of the transverse electric fields is smaller and most electrons are located within the region ($\eta_{x}\gg 0$, $|\eta_{\perp}|\sim0$), indicating the dominance of the longitudinal electric field acceleration over the transverse acceleration. Considering $0.5\lambda_0 \leq r \leq \lambda_0$, we see $|E_x|/|E_x|_{max}$ changes from 0.94 to 0.80, implying quasi-monoenergetic electron acceleration in our configuration.

Stable attosecond electron bunch formation requires simultaneous longitudinal acceleration and transverse focusing, i.e., $E_x<0$ and $E_r>0$. This demands $({2r^2}/{\sigma^2}-2)\cos\psi^{'}<0$ (note $kr^2/R\to 0$) and $\sin \psi^{'}>0$. As seen in Fig.~\ref{f2}(b), each attosecond electron bunch in our configuration are exactly positioned in the acceleration phase of the longitudinal electric field $E_x$. These bunches are relativistic and co-move with the laser pulse, so that they are quickly phase-locked longitudinally~\cite{Robinson13}. Figure~\ref{f2}(c) presents the radial field $E_r$ experienced by an attosecond bunch as marked in Fig.~\ref{f1}. Once phased-locked, the bunch always suffers positive electric field $E_r$, making the doughnut bunches close up after leaving the nanofiber. Figure~\ref{f2}(d) shows the electron phase-locking in a single laser period. For $r<\sigma_{0}$, $|E_x|$ is much stronger than for $r\gg\sigma_{0}$, which is sufficient to accelerate electrons to relativistic velocities in less than one laser cycle. Once entering the matched phase with positive $E_{r}$, these electrons are subject to a strong focusing force and are confined transversely. By contrast, as $r>\sigma_{0}$, $|E_x|$ gets much smaller so that the electron acceleration is weak and the phase-slip occurs, resulting in an increasing energy spread, even though the electrons are initially in the acceleration phase of the laser pulse. Thus, the appropriate phase for stable attosecond electron bunch formation satisfies $\psi^{'}\in[0,\pi/2]$ in one laser cycle.

When the energetic electrons are well confined near the laser axis, namely $-q_{e}(\vec{E} +\vec{v}\times\vec{B})_r\lesssim0$, we obtain the condition for the electron trapping, $(1-v_x/c)\sin\psi^{'}\gtrsim0$. This requires $\sin\psi^{'}\gtrsim0$ for each attosecond bunch, which is exactly the case here for the bunches in phase, so that the transverse electric and magnetic forces cancel each other out. However, in order to compensate the phase-slip during the acceleration (note $v_x<c$), the electron velocity should be high enough. Here the acceleration length approximates to the Rayleigh length, $L_a\sim\pi{\sigma _0}^2 /{\lambda}_0$. Thus $L_a(c-v_x)/c\le\lambda_0/4$ should be satisfied to keep electrons in phase, which requires
 \begin{eqnarray}
\frac{v_x}{c}\ge1-\frac{1}{4\pi}\Big(\frac{\lambda_0}{\sigma_0}\Big)^2.
\label{eq4}
 \end{eqnarray}
\noindent This implies an energy threshold of $\gamma_{th}\ge7.5$ in order to keep the electrons phased-locked in the LG laser fields.  For relativistic LG laser pulses, this is naturally satisfied, because such a laser pulse is capable of accelerating electrons to relativistic energies in less than one laser cycle. Once locked by the LG laser pulse, an electron acquires a velocity of $v_t=qE_x/m_e\times T_0$ in the first laser cycle, which should be such that $(1-v_t^2/c^2)^{-1/2}\ge\gamma_{th}$. This corresponds to a threshold $E_x/E_0>(\gamma_{th}^2-1)^{1/2}/(2\pi\gamma_{th})$. As seen in Eq.~(\ref{f2}) and Fig.~\ref{f2}(b) above, we take $E_x/E_0\sim(1/6)\times\sqrt{2e}a_0/(3\pi)$ to make sure most electrons in each bunch phase-locked. Therefore, we find that the laser amplitude threshold required for electron trapping in the LG laser field can be written as $a_0>9(\gamma_{th}^2-1)^{1/2}/(\sqrt{2e}\gamma_{th})=3.82$, if we ignore the multi-dimensional effects. By a series of 3D PIC simulations, we find the actual threshold is around $a_0=6$. Note that the LG laser pulses with such intensity are already available in several laboratories around the world~\cite{Leblanc17, Brabetz15}.

Figure~\ref{f3}(a) presents the electron trajectories up to $t=35T_{0}$, which illustrates clearly that most attosecond electrons originate from the left tip of the nanofiber. Since the laser electric field components $E_r$, $E_{\varphi}$, and $E_x$ are circularly-symmetric in the $yz$ plane, these electrons with the same $r$ experience similar acceleration. All these factors contribute to the quasi-monoenergetic electron acceleration as indicated in Fig.~\ref{f3}(b), where a peak energy of 375 MeV and a cutoff energy of 465 MeV are found at $t=60T_{0}$. The energy peak gradually increases in width, while the density is still above the critical density at $t=120T_0$ (400 fs). The final electron energy can be estimated as $\varepsilon_{k,max}=e\overline{E}_x L_a\approx470$ MeV, in excellent agreement with the simulations. Figure~\ref{f3}(c) shows the electron divergence angle distribution at $t=60T_0$. The energetic bunches obtained are distinctive with a small divergence angle of $<2^\circ$, a peak energy of $\sim$400 MeV and nC charges, which have diverse applications in various fields. For example, by applying the electron bunches obtained via our scheme in the recent Thomson scattering experiment~\cite{Yan17}, ultrabrilliant attosecond $\gamma$-ray bursts at the critical photon energy of 34 MeV are expected, since the beam charge here is enhanced by $10^3$ times while the beam energy spreads in both cases are comparable.
\begin{figure}[!ht]
\centering
\includegraphics[width=9.6cm]{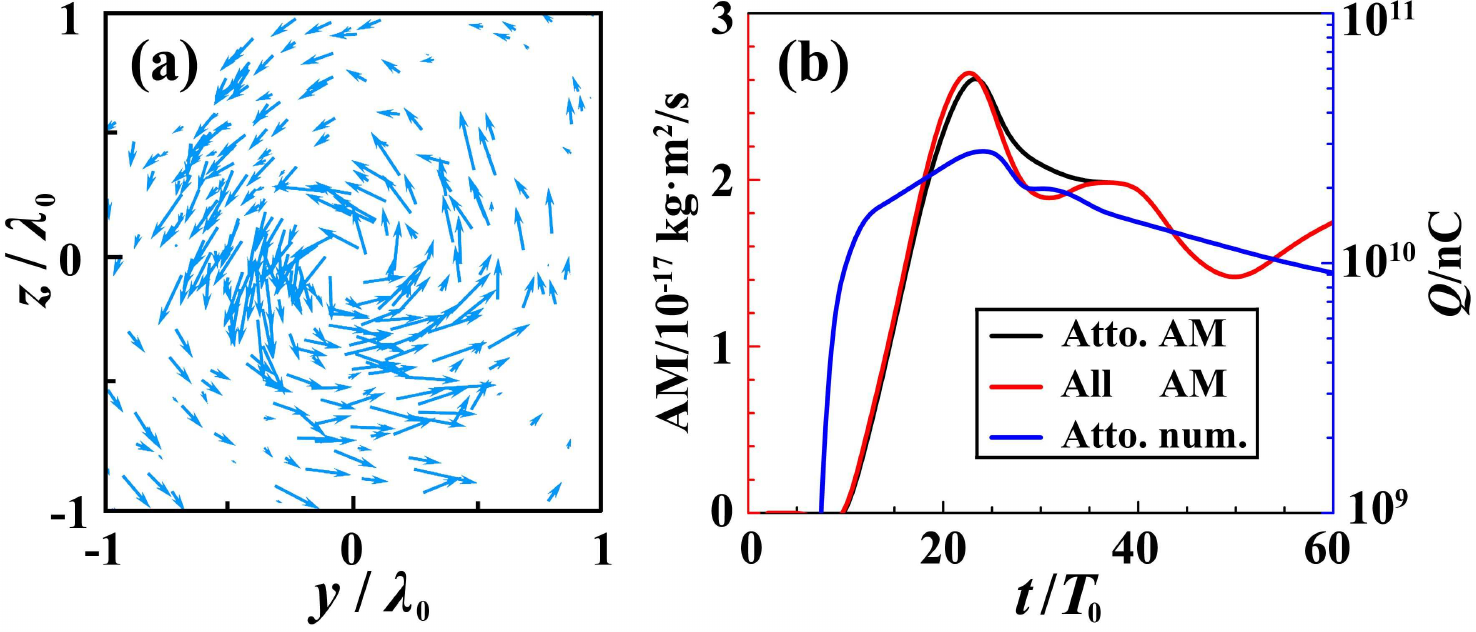}%
\caption{\label{f4}
Electron velocity field with AM larger than $2.73\times 10^{-28}$~kg$\cdot$m$^2$/s in the $yz$ plane with $x=44.1\lambda_0$ at $t=50T_{0}$ (a). Temporal evolution of the electron BAM and charge (b).}
\end{figure}

For a circularly-polarized LG laser pulse, its spin angular momentum (SAM) makes a charged particle spin with respect to its own axis, while its OAM is able to drive the particle rotate with respect to the laser axis. Under appropriate conditions, both can be transferred to particles, which is a fundamental question from a relativistic plasma optics perspective~\cite{Haines2001}. Here we employ a circularly polarized LG laser pulse with OAM but without total AM. Figure~\ref{f4}(a) exhibits the velocity field of electrons with AM larger than $2.73\times 10^{-28}$~kg$\cdot$m$^2$/s in the $yz$ plane. Apparently, most electrons rotate around the origin point and form a vortex. Figure~\ref{f4}(b) presents the evolution of the total electron BAM, where $\mathcal{L}=$$\sum_{i=1}^{n}[(y_{i}-y_{0})p_{z,i}-(z_{i}-z_{0})p_{y,i}]$ with $y_{0},z_{0}$ the target axis position in the $yz$ plane. A clear transfer of the laser OAM to the electron BAM is observed because of the circularly-symmetric non-zero azimuthal components of the LG laser field. Once in the phase-locking regime, these bunch electrons also suffer a fixed azimuthal electric field, tend to rotate with respect to the laser axis, and thus get AM from the laser pulse. As expected, almost all AM are carried away by the attosecond electron bunches. At $t=60T_0$, $\mathcal{L}$ is $>10^{-17}$ kg~m$^2$/s. This is a unique property of the bunches, which is otherwise absent in other cases using radially polarized normal Gaussian laser pulses~\cite{Karmakar2007,Salamin2007,Shell14,Zam2017} , making the bunches even more attractive for potential applications, e.g., generating x/$\gamma$-ray vortex via nonlinear Thomson scattering~\cite{Taira2017}.
\begin{figure}
\centering
\includegraphics[width=9.2cm]{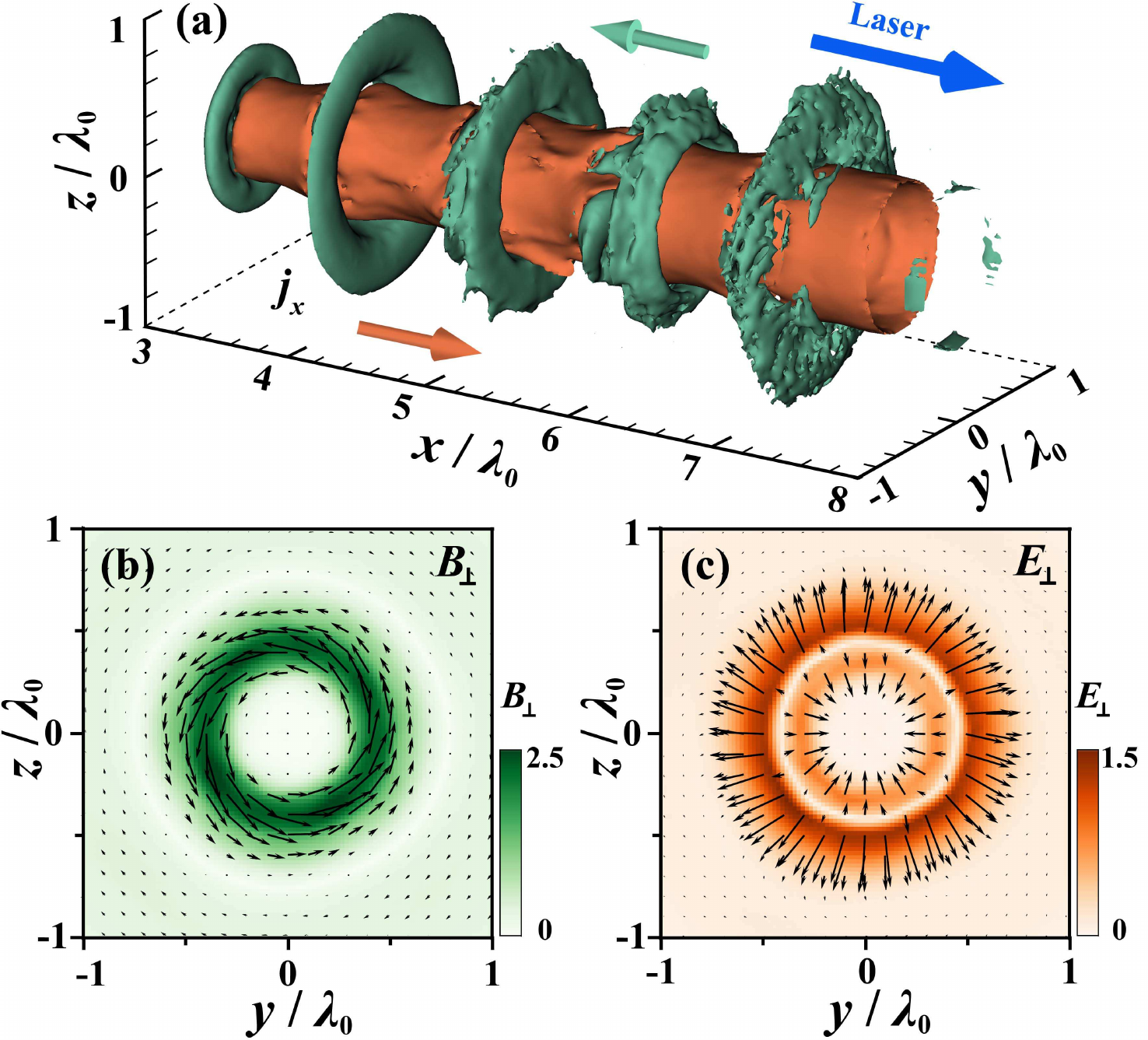}%
\caption{\label{f5}
3D isosurface of longitudinal current densities ($j_x=-1.1$~MA/$\upmu$m$^2$, green; $j_x^{'}=1.1$~MA/$\upmu$m$^2$, orange) at $t=10T_0$ (a). Averaged quasi-static magnetic field $B_{\perp}$ (b) and electric field $E_{\perp}$ (c). The electric and magnetic fields are respectively normalized by $E_0=m_ec\omega_0/q_e$ and $B_0=m_e\omega_0/q_e$.}
\end{figure}

\section*{Discussion}

The space charge effect plays an important role in determining the final electron number of each bunch. Here the transverse Coulomb field can be estimated as $E_{\perp}^{s}\sim(4\pi\epsilon_0)^{-1}(q_eN\gamma/{\sigma_0^2})$, where $\epsilon_{0}$ is the permittivity. Considering the relativistic cancellation of the magnetic and electric forces, the net transverse repulsion force approximates to $f_{\perp}^{s}\sim(4\pi\epsilon_{0})^{-1}q_e^2N/(\gamma\sigma_{0}^2)$. Since $f_{\perp}^{s} \propto 1/\gamma$ is much smaller than the Lorentz force as $\gamma\gg1$, the repulsion force is suppressed significantly, so that we can take into account only the laser force here. The electron number in each bunch is thus limited to $N_{e,max}<(\sigma_{0}/r_{e})[p_{\perp}/(m_{e0}c)]^2$, according to the impulse theorem $\Delta p_{\perp}\sim (m_{e0}\gamma\sigma_{0}/p_{\perp})<{p_{\perp}}f_{\perp}^{s}$~\cite{Stupakov01}. Here, $r_{e}=(4\pi \epsilon_0)^{-1}{e^2}/({m_e c^2})$ is the classic electron radius. As electrons are dragged out, $f_{\perp}^{s}$ increases accordingly. Once $f_{\perp}^{s}$ approaches the maximum, electrons are no longer dragged out. Finally, we find the total electron number is up to $10^{10}$ in our simulations, about $\sim61.3\%$ of which originate from the left tip. This implies that we may replace the nanofibre with micro-droplets to facilitate experiments. Note that it is easier to produce micro-droplets in large numbers and consistent with high repetition rate lasers~\cite{droplet}. One can thus test the set-up in experiments with the current and forthcoming LG-mode lasers in the near future.
\begin{figure}
\centering
\includegraphics[width=9.4cm]{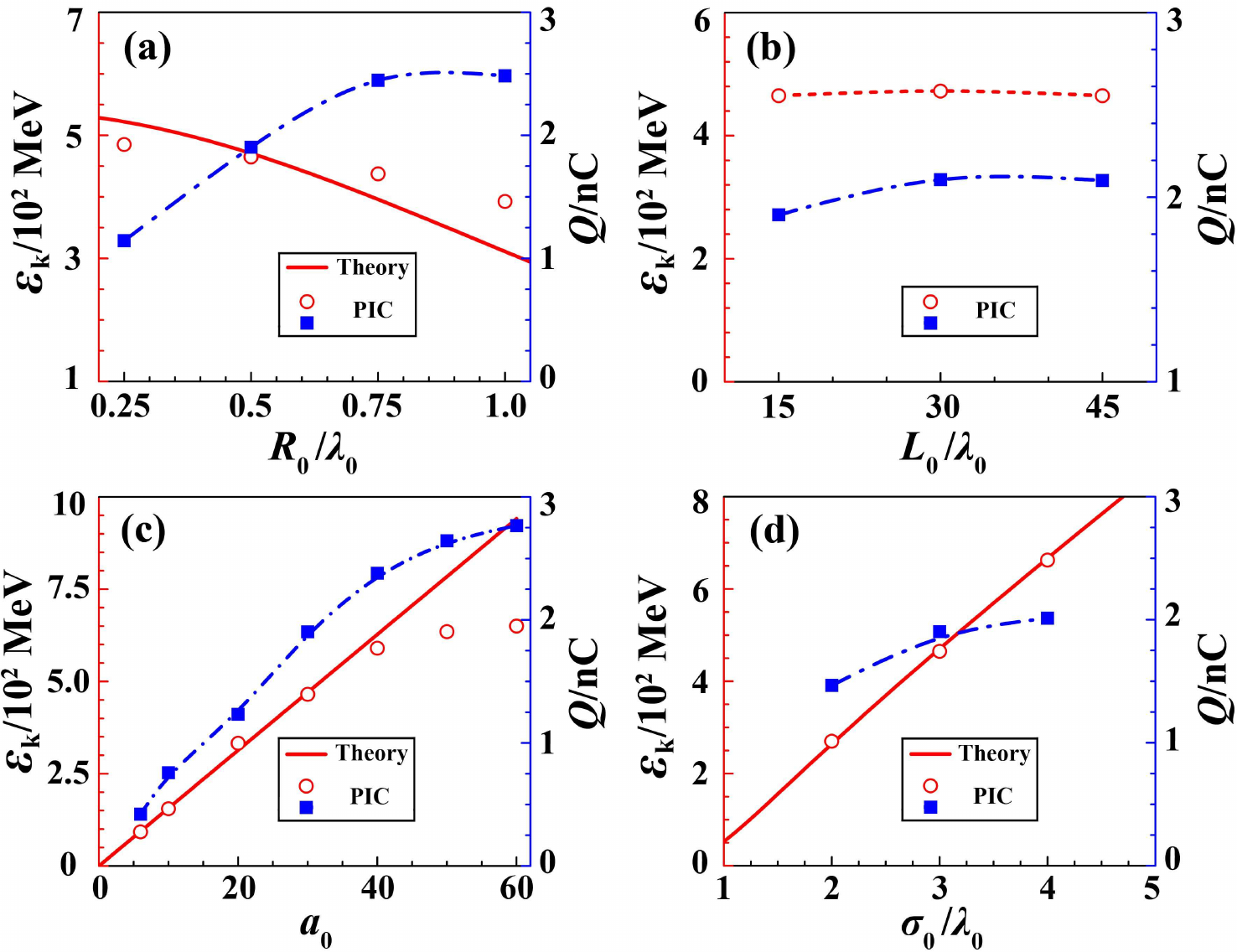}%
\caption{\label{f6}
Parametric influence of the fiber radius $R_0$ (a) and length $L_{0}$ (b), the laser dimensionless parameter $a_0$ (c) and focal size $\sigma_0$ (d) on the achieved bunch energy and charge.}
\end{figure}

Figure~\ref{f5}(a) shows the 3D isosurface of the longitudinal current density at $t=10T_0$. Since abundant electrons are dragged out of the nanofiber, strong charge separation fields also form in the nanofiber, resulting in a strong return current to satisfy the plasma neutralization condition~\cite{Kaymak16}. The return current tends to pump electrons to the left tip, some of which can be distinguished from the trajectories above. During the process, quasi-static electromagnetic fields around the nanofiber are also induced, as seen in Fig.~\ref{f5}(b) and (c), though their amplitudes are much smaller as compared to the laser fields.

Finally, we explore the influence of the laser and target parameters on the bunch generation, as summarized in Fig.~\ref{f6}. Here, we vary a single parameter in each case and keep all other parameters fixed. By increasing the fiber radius, the longitudinal electric field $E_x$ decreases and the phase-slip occurs more quickly, so that the electron energy decreases. Correspondingly, the radial electric field $E_r$ increases and more surface electrons are dragged out, and finally the total beam charge increases. In experiments, the pre-pulse may expand the front end of the nanofiber so that the real diameter increases. However, once the diameter becomes larger than a critical value ($r_{th}=\sigma_0$), the longitudinal electric field $E_x$ in Eq.~(\ref{eq2}) becomes positive, which accelerates surface electrons along the -x direction (see Supplementary Fig. S2 online). As a result, these electrons move backward and are separated from the forward bunches. As discussed above, since most bunch electrons come from the front tip, the beam charge and energy weakly depend on the fiber length, which is demonstrated by simulations in Fig.~\ref{f6}(b). Meanwhile, the laser intensity plays an important role in the final beam energy and flux. Figure~\ref{f6}(c) indicates that, by increasing the laser dimensionless parameter $a_0$,  the longitudinal electric field $E_x$ rises linearly, leading to higher electron energies, larger beam flux, and longer pulse lifetime ($>500$ fs). The regime is also valid for a much lower laser intensity and an example simulation with $a_0=10$ is given in Supplement. On the other hand, using a larger laser focal spot, the acceleration length of the beam increases and much higher electron energies are achieved. Other experimental factors on attosecond bunch formation are also discussed (see see Supplementary Fig. S3 and S4 online). Because there exists a laser intensity threshold for the short bunch formation in our scheme, one may also tailor the laser pulse to realize a single attosecond electron bunch generation.

In summary, we report a novel efficient scheme for generating a stable and dense attosecond electron bunch train by irradiating a nanofiber target with an circularly polarized, intense LG-mode laser pulse. The generated electron bunch train results from the unique field structure of the circularly polarized LG laser pulse, and is distinctive with ultrashort bunch duration, high beam charge and density, small divergence angle, and high BAM. These features are extremely attractive for generating ultrabrilliant attosecond x/$\gamma$-ray bursts, enhancing coherent ionization losses~\cite{CIL2001}, driving plasma-based electron acceleration, and generating high-energy-density states of matters. We also observe a clear transfer of the laser OAM to the attosecond electron BAM, offering a new avenue towards the generation of bright x/$\gamma$-ray vortex via nonlinear Thomson scattering~\cite{Taira2017}.

\section*{Methods}
\subsection*{Numerical modeling.}
We first carry out a series of 3D simulations to investigate the physics of the LG-mode laser interaction with a nanofiber target. The PIC code Virtual Laser Plasma Lab (VLPL~\cite{Pukhov99,Yu2014}) is employed, which allows direct fully electromagnetic simulations of relativistic laser plasma interactions. In initial simulations, a hydrogen nanofiber with plasma density of $n_{e}=20n_c$ is used. Here, $n_c$=1.12$\times10^{21}~$cm$^{-3}$ is the critical density for a $\lambda_{0}=1~\upmu$m laser. The target could be solid-hydrogen, hydrocarbon or, carbon and silicon materials in experiments~\cite{Jiang16}. The fiber radius and length are $R_0=0.5~\upmu$m and $L_0=15~\upmu$m, respectively. The particle number per cell (PPC) is 27 for both electrons and ions. Absorbing boundary conditions for the fields and particles are employed in the $x$ direction (the laser propagation direction), while periodic ones for the fields and absorbing ones for the particles are used in the $y$ and $z$ directions. The initial electron and ion temperatures are assumed to be 1 keV and 0.01 keV, respectively. A left-hand circularly-polarized LG laser pulse with mode ($\ell,~p$) is focused on the left side of the fiber and propagates along the $x$-axis. Here the integer $\ell$ denotes the number of azimuthal phase cycles and $(p+1)$ represents the number of radial nodes. The laser electric field is given by
\begin{eqnarray}
\begin{aligned}
\vec{E}_{\perp}\big(LG_p^{\ell}\big)=C_p^{\ell}  E_{L0}(-1)^p e^{-\frac{r^2}{\sigma ^2}}\Big(\frac{\sqrt{2}r}{\sigma}\Big)^{\ell}L_p^{\ell}\Big(\frac{2r^2}{\sigma ^2}\Big)g\Big(x-ct\Big)
\Big(\sin \psi \vec{e}_y -  \cos \psi  \vec{e}_z\Big),\label{eq0}
\end{aligned}
\end{eqnarray}

\noindent where $E_{L0}$=$m_{e0}c\omega_{0}a_{0}/{q_e}$ is the peak amplitude of the laser electric field with $m_{e0}$ the electron rest mass, $c$ the speed of light in vacuum, $\omega_{0}$ the laser angular frequency, $q_e$ the charge unit, and $a_{0}$ the laser dimensionless parameter, $r=({y^2+z^2})^{1/2}$, $\sigma = \sigma_0 (1+ x^2/ f^2)^{1/2}$ with $\sigma_0=3\lambda_{0}$ the focal spot radius, 
and $f=\pi \sigma_0 ^2/\lambda_0$ the Rayleigh length, $g(x-ct)=\cos^2[\pi (x-ct)/(2c\tau)]$ for $-c\tau \le x-ct\le c\tau$ with $\tau=5T_0$ and $T_0=\lambda_0/c$, and $L_p^{\ell}$ is the generalized Laguerre polynomial. Here, $\psi$=$ -kx+\omega_0 t-kr^2/(2R)-(\ell+2p+1)\arctan(x/f)+ \ell \varphi $ denotes the phase term with $k$ the wave number, $R= x + f^2/x$ the curvature radius of the wave-front, $(\ell+2p+1)\arctan (x/f)$ the Guoy phase of the mode, and $\varphi$ the azimuthal angle. We use the ($1,~0$) mode of the LG laser, which implies an azimuthal phase term $e^{i \varphi}$ and a spiral wave-front of the laser pulse.

\subsection*{Data availability.} Data associated with research published in this paper can be accessed at http://dx.doi.org/10.15129/f9d7cc35-8f3c-461c-9bbe-ed6a9c8a667b.

\section*{Acknowledgements}

The authors acknowledge encouraging discussions with A. Pukhov, C. Baumann and M. Y. Yu. This work was financially supported by the NSFC (Grant Nos. 11622547, 11675264, 11474360, 11375265, 11475030, 11705280); National Basic Research Program of China (Grant No. 2013CBA01504); Science Challenge Project (Grant No. TZ2016005); Hunan Provincial Natural Science Foundation of China (Grant No. 2017JJ1003); Hunan Provincial Science and Technology Program (Grant No. 2017RS3042); Research Project of NUDT (JQ14-02-02); EPSRC (Grant No. EP/M018091/1, EP/R006202/1 and EP/N028694/1); a Leverhulme Trust Grant at Strathclyde.

\section*{Author contributions}

L.X.H. conducted the numercical simulations, analyzed the data, and wrote the initial manuscript. T.P.Y conceived the idea, developed the theoretical model, and organised the manuscript. Z.M.S., J.V. and P.M. clarified the underlying physics and contributed to the wirting. D.B.Z., Y.Y., P.M., and F.Q.S. discussed the results and revised the manuscript.

\section*{Additional information}

\textbf{Competing interests:} The authors declare that they have no competing interests.

\end{document}